\documentclass[prd, twocolumn, nofootinbib, floatfix]{revtex4-1}

\usepackage{amsmath}
\usepackage{graphicx}
\usepackage{dcolumn}
\usepackage{bm}
\usepackage{epsfig}
\usepackage{amssymb,latexsym,mathrsfs}
\usepackage{graphicx}
\usepackage{color}

\newcommand{\be}{\begin{equation}}
\newcommand{\ee}{\end{equation}}
\newcommand{\bs}{\begin{split}} 
\newcommand{\bea}{\begin{eqnarray}}
\newcommand{\eea}{\end{eqnarray}}

\newcommand{\om}{\Omega_m}
\newcommand{\op}{\Omega_\phi}

\begin{document}

\title{Purely Kinetic Coupled Gravity} 
\author{Giulia Gubitosi$^1$, Eric V.\ Linder$^{1,2}$} 
\affiliation{$^1$ Berkeley Lab \& University of California, Berkeley, 
CA 94720, USA}
\affiliation{$^2$ Institute for the Early Universe WCU, Ewha Womans 
University, Seoul, Korea}

\begin{abstract}
Cosmic acceleration can be achieved not only with a sufficiently flat 
scalar field potential but through kinetic terms coupled to gravity. 
These derivative couplings impose a shift symmetry on the scalar field, 
aiding naturalness.  We write the most general purely kinetic action 
not exceeding mass dimension six and obeying second order field equations. 
The result reduces to a simple form involving 
a coupling of the Einstein tensor with the kinetic term 
and can be interpreted as adding a new term to Galileon 
gravity in curved spacetime.  We examine the cosmological implications 
of the effective dark energy and classify the dynamical attractor solutions, 
finding a quasistable loitering phase mimicking late time acceleration by 
a cosmological constant. 
\end{abstract}

\date{\today} 

\maketitle

\section{Introduction} 

Modifications to the Einstein-Hilbert action for gravity can provide 
mechanisms for explaining cosmic acceleration, either the inflationary 
epoch in the early universe or the present dark energy era.  These 
often give additional degrees of freedom which can be viewed as scalar 
fields.  However, in the presence of a scalar field potential there is 
no particular need to modify the action; a canonical, minimally coupled 
scalar field can induce acceleration by itself.  It is of interest, 
therefore, to consider purely kinetic, i.e.\ free field, theories where 
a noncanonical kinetic term or coupling to the metric is responsible for 
the physics of acceleration. 

Recently, great interest has arisen in theories where the scalar field 
has a shift symmetry or a Galilean symmetry.  These protect the theory 
against high energy radiative corrections, reducing the unnaturalness 
of rigidly translating a high energy Lagrangian into the same form at 
the drastically lower energies of the present universe.  Such Galileon 
theories \cite{galileon,galileondeffayet} indeed have noncanonical kinetic 
terms and sometimes coupling to metric quantities.  Noncanonical kinetic 
models without coupling, such as k-essence and k-inflation theories 
\cite{armend1,armend2}, have also been successful at giving rise to 
cosmic acceleration and particularly inflationary scenarios.  

One of the problems with adding a scalar degree of freedom, particularly 
with noncanonical kinetic terms and gravitational coupling, is 
the appearance of higher order field equations that have an ill defined 
initial value problem and the presence of ghosts.  Interesting early work 
on this was carried out by \cite{amend9302010,bekenstein}.  Here we revisit 
this from a different angle, coming to similar conclusions as a number 
of papers, e.g.\ \cite{9901051,capoz9906051,sushkov,kehagias}, that a 
particular form of the coupling is essential.  While these papers 
examine the impact on inflation, in this article we address late time 
cosmic acceleration, use a purely kinetic theory with shift symmetry, 
include matter fields, and calculate the dynamics of the effective dark 
energy.  In particular we identify a ``loitering'' effective cosmological 
constant epoch and discuss extensions to the Galileon framework. 

In Sec.~\ref{sec:action} we consider the most general action involving 
derivative couplings and impose conditions on the order of the field 
equations and the mass dimension of the terms, giving a unique form.  
The equations of motion for the resulting action are derived in 
Sec.~\ref{sec:eom} and specialized to a cosmological background.  We 
solve the equations and identify dynamical attractor solutions in 
Sec.~\ref{sec:cos} and discuss the cosmological implications for the 
cosmic expansion and acceleration. 

We will adopt units such that $\hbar=c=1$.

\section{Action with Derivative Couplings} \label{sec:action} 

Scalar-tensor theories generally involve coupling of the field itself 
to the Ricci curvature appearing in the Einstein-Hilbert action, changing 
$R/G_N$ to $F(\phi)\,R$, where $G_N$ is Newton's constant for the 
gravitational coupling and $\phi$ is the scalar field value.  Derivative 
couplings instead involve the covariant derivative $\phi_\mu\equiv 
\nabla_\mu\phi$ coupled to functions of the metric $g_{\mu\nu}$ in such 
a way to give a scalar quantity in the action. 

Such derivative couplings arise in string theory and higher dimensional 
or massive gravity (see, e.g., 
\cite{gross,amend9302010,galileondeffayet,kehagias} and references 
therein).  One attractive motivation for 
considering derivative couplings is the idea that when the field freezes, 
e.g.\ in a de Sitter asymptotic state, then the couplings turn off and 
general relativity is restored.  Such theories may in fact have an 
attractor toward an accelerating universe that looks like standard 
cosmological constant $\Lambda$CDM.  Another motivation is that mentioned 
in the Introduction, that the theory is at least somewhat protected 
and natural.  See \cite{amend9302010,capoz9906051} for some early work 
on kinetic coupling. 

The most general action coupling functions of the metric with field 
derivatives, not exceeding mass dimension six, is 
\be 
\begin{split}
S_{c}&=\int d^{4}x\sqrt{-g}\,\left[ R\,\left( \frac{c_1}{M_{p}^{2}} \phi_{\mu}\phi_{\nu}g^{\mu\nu} 
+\frac{c_2}{M_{p}}\square\phi\right)+\frac{c_3}{M_{p}^{2}} R^{\mu\nu}\phi_{\mu}\phi_{\nu}\right.\\ 
&\qquad +\left. 
\frac{c_4}{M_{p}^{2}} R^{\alpha\beta\gamma\delta}\Phi_{\alpha\beta\gamma\delta}\right] \,. 
\label{eq:action} 
\end{split} 
\ee 
where the $c_{i}$ are dimensionless coefficients. 
Note that this is in addition to the normal Einstein-Hilbert action and 
matter Lagrangian, i.e.\ we concentrate in this section on the difference. 
For simplicity we use the notation $\phi_\mu=\nabla_\mu\phi$ and 
$\Phi_{\alpha\beta\gamma\delta}$ is a function of the $\phi_\mu$ discussed 
below. 

One of the interesting aspects of this action is that it is somewhat 
reminiscent of a Gauss-Bonnet action, $g_1 R\cdot R+g_2 R_{\mu\nu} R^{\mu\nu} 
+g_3 R_{abcd}R^{abcd}$.  However here, instead of coupling gravity to 
itself we are coupling gravity to the scalar field derivatives.  

Terms that do not appear in Eq.~(\ref{eq:action}) include ones like 
$\phi\square\phi$  -- this is equivalent to the $c_1$ term 
up to a total derivative when avoiding more than two derivatives of 
any quantity, and in any case does not obey shift symmetry -- and ones 
involving more than two products of $\phi_\mu$ (or more than one 
$\phi_{\mu\nu}$) since these will exceed mass dimension six and so be 
suppressed by higher powers of the Planck mass. 

The quantity $\Phi_{\alpha\beta\gamma\delta}$ must have the same symmetry 
properties as $R^{\alpha\beta\gamma\delta}$ in order to give a 
contribution.  There is only one possibility not exceeding mass dimension six: 
\begin{equation}
\Phi_{\alpha\beta\gamma\delta}=\phi_{\alpha}\phi_{\gamma} 
g_{\beta\delta} -\phi_{\beta}\phi_{\gamma}g_{\alpha\delta} -\phi_{\alpha} 
\phi_{\delta} g_{\beta\gamma}+\phi_{\beta}\phi_{\delta}g_{\alpha\gamma}\,, 
\end{equation} 
but it contracts against $R^{\alpha\beta\gamma\delta}$ to give 
$4\phi_\alpha\phi_\gamma R^{\alpha\gamma}$ and so it is absorbed into the 
$c_3$ term.  

Another term that would have the correct symmetry properties is
\begin{equation}
\Phi_{\alpha\beta\gamma\delta}=\phi_{\alpha\gamma}\phi_{\beta\delta} 
-\phi_{\beta\gamma}\phi_{\alpha\delta} \,, 
\end{equation} 
or also a $\Phi_{\alpha\beta\gamma\delta}$ involving terms of the kind 
$\phi_\alpha\phi_\gamma\phi_{\beta\delta}$.  But the first kind of  
$\Phi_{\alpha\beta\gamma\delta}$ has mass dimension 8 and the other one, 
besides being more that quadratic in the fields, has mass dimension 9.  
Even if one wanted to consider these terms, one would find that they would 
be forced to vanish due to causing more than second order field equations, 
without canceling against any other term. 

So we have no $c_{4}$ terms allowed into $S_{c}$. 

To find out the conditions on the coefficients $c_{1}$, $c_{2}$, $c_{3}$ 
that avoid the appearance of higher than second order derivatives we have 
to look at the variation of the action $\delta S_{c}/\delta g^{\mu\nu}$. 
\begin{eqnarray}
\delta S_{c}&=&\int d^{4}x\sqrt{-g}\Big[ -\frac{1}{2}g_{\mu\nu}\delta g^{\mu\nu} R\left( \frac{c_1}{M_{p}^{2}} \phi_{\alpha}\phi_{\beta}g^{\alpha\beta}+ \frac{c_2}{M_{p}}\square\phi\right)\nonumber \\
&+&\delta R \left( \frac{c_1}{M_{p}^{2}} \phi_{\alpha}\phi_{\beta}g^{\alpha\beta}+ \frac{c_2}{M_{p}}\square\phi\right)+ \frac{c_2}{M_{p}} R \delta(\phi_{\mu\nu})g^{\mu\nu}\nonumber \\
&+&R \left( \frac{c_1}{M_{p}^{2}} \phi_{\mu}\phi_{\nu}  + \frac{c_2}{M_{p}}\phi_{\mu\nu}\right)\delta g^{\mu\nu}-\frac{1}{2} \frac{c_3}{M_{p}^{2}}g_{\mu\nu}\delta g^{\mu\nu} R^{\alpha\beta}\phi_{\alpha}\phi_{\beta}\nonumber\\
&+& \frac{c_3}{M_{p}^{2}}\delta(R_{\alpha\beta})\phi^{\alpha}\phi^{\beta}+ \frac{c_3}{M_{p}^{2}}R_{\alpha\beta}\phi_{\mu}\phi_{\nu}\delta (g^{\mu\alpha}g^{\nu\beta})\Big]\label{eq:deltaSc}
\end{eqnarray} 

Exploiting the following formulas for the variations of some important 
quantities: 
\begin{eqnarray}
&&\delta R\equiv \delta(R_{\mu\nu})g^{\mu\nu}+ R_{\mu\nu}\delta(g^{\mu\nu})\nonumber\\
&&\delta(R_{\mu\nu})=(\delta \Gamma^{\alpha}\,_{\mu\nu})_{;\alpha}-(\delta \Gamma^{\alpha}\,_{\mu\alpha})_{;\nu}\nonumber\\
&&\delta(R_{\alpha\beta\gamma\delta})=\delta (g_{\alpha\rho} )R^{\rho}\,_{\beta\gamma\delta}+g_{\alpha\rho}\left[(\delta\Gamma^{\rho}\,_{\beta\delta})_{;\gamma}-(\delta\Gamma^{\rho}\,_{\beta\gamma})_{;\delta} \right]\nonumber\\
&&\delta(\nabla_{\mu}\nabla_{\nu}\phi)=\delta(\Gamma^{\alpha}\,_{\mu\nu})\partial_{\alpha}\phi\nonumber\\
&&\delta\Gamma^{\kappa}\,_{\sigma\tau}=\frac{1}{2}g^{\kappa\rho}\left[(\delta g_{\rho\sigma})_{;\tau}+(\delta g_{\rho\tau})_{;\sigma}-(\delta g_{\sigma\tau})_{;\rho}\right] \label{eq:UsefulFormulas}
\end{eqnarray} 
one can see that the terms that could potentially lead to higher 
order derivatives in the equations of motion are: 
\begin{equation}
\sqrt{-g}\delta( R_{\mu\nu})\left[g^{\mu\nu} \left( \frac{c_1}{M_{p}^{2}} \phi_{\alpha}\phi_{\beta}g^{\alpha\beta}+ \frac{c_2}{M_{p}}\square\phi\right)+ \frac{c_3}{M_{p}^{2}}\phi^{\mu}\phi^{\nu}
\right] \,. 
\end{equation} 
Writing explicitly the variations and integrating by parts it can be shown 
that the terms proportional to $c_{2}$ do not vanish and are the only ones 
producing fourth order derivatives in the equations of motion, so that 
they cannot cancel out with anything. Thus we have to ask that $c_{2}=0$. 

As regards the other terms, proportional to $c_{1}$ and $c_{3}$, they 
generate third order derivatives in the equations of motion, giving a 
contribution proportional to 
\begin{eqnarray}
&&-\delta g_{\gamma\lambda}g^{\gamma\lambda}
\bigg[2 c_{1}  (\phi_{\mu\beta}\,^{\beta}\phi^{\mu})+ \frac{c_{3}}{2} (\phi_{\mu\beta}\,^{\beta}\phi^{\mu}+\phi^{\beta}\,_{\beta\mu}\phi^{\mu})\bigg]\nonumber\\ 
&&\quad +\delta g_{\gamma\lambda}\bigg[2 c_{1} \,\phi_{\mu}\,^{\gamma\lambda}\phi^{\mu}+c_{3}  \phi^{\lambda\gamma}\,_{\mu}\phi^{\mu}\bigg]\nonumber\\
&=& -\delta g_{\gamma\lambda}g^{\gamma\lambda}
\bigg[(2 c_{1}+c_{3})  (\phi_{\mu\beta}\,^{\beta}\phi^{\mu})+\frac{c_{3}}{2}g^{\beta\alpha}R^{\rho}\,_{\alpha\beta\mu}\phi_{\rho}\phi^{\mu})\bigg]\nonumber\\&&+\delta g_{\gamma\lambda}\bigg[(2 c_{1}+c_{3}) \,\phi_{\mu}\,^{\gamma\lambda}\phi^{\mu}+c_{3}  g^{\lambda\alpha}g^{\gamma\beta}R^{\rho}\,_{\beta\alpha\mu}\phi_{\rho}\phi^{\mu}\bigg] \,. 
\label{eq:allCovariantDerivs}
\end{eqnarray}
In order not to have third order derivatives in the equations of motion 
we require $2c_{1}+c_{3}=0$.  This condition has been recognized 
previously, e.g.\ \cite{9901051,capoz9906051,sushkov,kehagias}, although 
generally starting from different approaches and with different applications. 

Thus our conclusion is that to have no more than second order field 
equations one must have $c_1=-c_3/2$, $c_2=0=c_4$.  This puts the initial 
action Eq.~(\ref{eq:action}) into the final form 
\be 
\begin{split} 
S_{c}&=\frac{ c_3}{M_{p}^{2}}\int d^4 x\sqrt{-g}\,\left[-\frac{1}{2}R g^{\mu\nu}+R^{\mu\nu}\right] 
\phi_\mu\phi_\nu\\ 
&=\frac{ c_3}{M_{p}^{2}}\int d^4 x\sqrt{-g}\,G^{\mu\nu}\phi_\mu\phi_\nu \,, 
\end{split} 
\ee 
where $G^{\mu\nu}$ is the Einstein tensor.  

The simplicity of this result is interesting.  Such an action has been 
considered before, e.g.\ from disformal field theories \cite{kaloper} 
and Higgs inflation \cite{kehagias}, and applied to inflation 
\cite{sushkov} and the cosmological constant \cite{copeland}, but we have 
come to it from a different direction.  

This additional contribution to the action vanishes in flat space, so 
it does (trivially) obey Galilean symmetry in flat space as well as 
shift symmetry.  In curved space, Galilean symmetry is broken as for 
the standard Galileon theories.  Oddly, this term is not included in 
the usual treatments of Galileon gravity \cite{galileon}.  We 
suspect the reason is that in generalizing the flat spacetime Galileon 
theory to curved spacetime \cite{galileondeffayet} the focus was on 
the ${\mathcal L}_4$ and ${\mathcal L}_5$ Galileon terms (of mass dimension 
10 and 13 respectively), since only these could generate higher than 
second order derivatives in the field equations.  Thus our term, which 
is like a covariantization of ${\mathcal L}_2$ in curved spacetime, was 
not considered.  It does appear to be an interesting and healthy extension 
to general relativity, worthy of further exploration. 

Another view is to consider the field derivative as similar to a 
vector field, tying this to vector gravity theories.  This analogy 
has been investigated by \cite{daniel} for the case of compact objects 
rather than cosmology, and using a coupling to $R^{\mu\nu}$ rather 
than $G^{\mu\nu}$.

\section{Equations of Motion} \label{sec:eom}

The final form of the action in the purely kinetic coupling theory is 
the sum of the canonical action and the coupling action, 
\be 
S= \int d^4 x\sqrt{-g}\,\left[\frac{M_P^2}{2}R -\frac{1}{2}g^{\mu\nu} 
\phi_\mu\phi_{\nu}+\frac{c_3}{M_P^2}\,G^{\mu\nu}\phi_\mu\phi_\nu 
-{\mathcal{L}_m}\right]\,, \label{eq:fullaction} 
\ee 
where $M_P^2=1/(8\pi G)$ is the reduced Planck mass and ${\mathcal L}_m$ 
is the matter Lagrangian.

\subsection{General Considerations and Implications} 

Note that the kinetic term in the Lagrangian was forced by the condition 
of second order field equations to a very special form, one with interesting 
implications.  Let us write a general separable Lagrangian 
\be 
{\mathcal L}=F(\phi_\mu\phi_\nu)\,E(g^{\mu\nu}) \,, 
\ee 
where proper contraction of indices is understood.  Then upon variation 
with respect to the field $\phi$ we obtain 
\be 
\bs 
0&=\nabla_\mu\left(E\frac{\partial F}{\partial X}\nabla_\nu\phi\right)\\ 
&=(\nabla_\mu E)\, F_X \nabla_\nu\phi +E\,F_X\nabla_\mu\nabla_\nu\phi 
+E\,F_{XX} (\nabla_\mu\nabla_\mu\phi)\, \nabla_\nu\phi \nabla_\nu\phi \,. 
\end{split}
\ee 
Here $X=(1/2)g^{\mu\nu}\phi_\mu\phi_\nu$ is the canonical kinetic energy 
and $F_X=dF/dX$. 

If there is no coupling, so $E=1$, then we have the k-essence modified 
Klein-Gordon equation.  See \cite{armend1,armend2,depl0705} for details 
of k-essence, its accelerating cosmological solutions, and the purely 
kinetic k-essence behavior.  

However, if we have $X$ (really $\phi_\mu\phi_\nu$) appearing linearly 
in the action, as required from our mass dimension constraint, then 
$F_X=1$, $F_{XX}=0$ and the last term in the equation of 
motion vanishes.  But in fact the first term also vanishes due to our 
requirement of second order field equations; the action was forced to 
a particular form of 
\be 
E^{\mu\nu}= -\frac{1}{2}g^{\mu\nu}+\frac{c_3}{M_P^2}\,G^{\mu\nu} \,, \label{eq:edef} 
\ee 
and both $g^{\mu\nu}$ and $G^{\mu\nu}$ are covariantly conserved so 
that $\nabla_\mu E^{\mu\nu}=0$.  Thus, the form of the Klein-Gordon 
equation is 
\be 
E^{\mu\nu}\,\nabla_\mu\nabla_\nu\phi=0 \,. \label{eq:ekg} 
\ee 

This resembles the unmodified form of the Klein-Gordon equation if one 
use a deformed metric $\tilde g^{\mu\nu}=g^{\mu\nu}-(2c_3/M_P^2)G^{\mu\nu}$ 
(such a transformation is sometimes called disformal, to distinguish it 
from conformal transformations, and has interesting implications 
\cite{bekenstein,kaloper}) to contract the indices (but not in the 
derivatives).  One can turn this 
around and say that if one insists that the form of the field equation of 
motion appears ``conserved'' in this way under such a deformed metric then 
one is led to the purely kinetic coupling action of 
Eq.~(\ref{eq:fullaction}).  That is, the special form involving only the 
combination $G^{\mu\nu}$ ensures that the deformed metric remains covariantly 
conserved.  Also note that the kinetic coupling still induces modifications 
through the contribution to the expansion history $H$ within the covariant 
derivatives. 

A final point of interest is that another solution exists where 
$E^{\mu\nu}=0$.  This implies 
\be 
g^{\mu\nu}-\frac{2c_3}{M_P^2}\,G^{\mu\nu}=0\,, 
\ee 
which has a solution in de Sitter space.  This determines the coupling 
constant 
\be 
c_3=-\frac{M_P^2}{6H^2} \qquad ({\rm de\ Sitter})\,. \label{eq:cdes} 
\ee 
Such a solution is quite interesting as it indicates that a possible 
fixed point exists for the theory giving an accelerating universe 
asymptotically approaching a cosmological constant state -- without 
any explicit cosmological constant, or even potential, in the theory. 
We discuss this further after we derive the full equations of motion.

\subsection{Field Equations} 

The dynamics of the system described by the action of 
Eq.~(\ref{eq:fullaction}) is found by minimizing the action with respect 
to variations of the metric and variations of the field.  Variation with 
respect to the metric gives the spacetime dynamics, i.e.\ the expansion 
as a function of time, while variation with respect to the field gives 
the evolution equation for the field. Of course, since the action 
(\ref{eq:fullaction}) couples the metric and the field, we will have a 
set of coupled differential equations, describing the interplay between 
the expansion evolution and the field dynamics.  Some dynamical analysis, 
focused on de Sitter or superacceleration states, has been touched on 
in \cite{9901051,10125719}. 

The variation of the action with respect to the metric can be found 
by exploiting the relations listed in Eq.~(\ref{eq:UsefulFormulas}), 
and integration by parts as needed.  Requiring that the variation 
vanishes, $\delta S/\delta g^{\mu\nu}=0$, the dynamical equations for 
the metric are 
\begin{eqnarray}
&&\frac{M_{P}^{2}}{2}(R_{\mu\nu}-\frac{1}{2}g_{\mu\nu}R)-\frac{1}{2}\phi_{\mu}\phi_{\nu}\nonumber\\
&&+\frac{1}{4}g_{\mu\nu}\partial_{\alpha}\phi\partial_{\beta}\phi g^{\alpha\beta}  -\frac{c_{3}}{2M_{P}^{2}}\Big[G_{\mu\nu}\, \phi_{\alpha}\phi_{\beta}g^{\alpha\beta} \nonumber\\
&&+R \phi_{\mu}\phi_{\nu} +g_{\mu\nu} R^{\alpha\beta}\phi_{\alpha}\phi_{\beta}-4R_{\nu\beta}\phi_{\mu}\phi_{\alpha}g^{\alpha\beta}\nonumber\\
&&-2\phi_{\lambda\mu}\phi^{\lambda}\,_{\nu}+2R^{\alpha}\,_{\nu\mu\beta}\phi_{\alpha}\phi^{\beta}+2\phi_{\nu\mu}\phi^{\kappa}\,_{\kappa} \nonumber\\
&&+g_{\mu\nu}\, \phi_{\alpha\beta}\phi^{\alpha\beta}-g_{\mu\nu}\phi^{\kappa}  g^{\sigma\tau}R^{\rho}\,_{\tau\sigma\kappa}\phi_{\rho}\nonumber\\
&&-g_{\mu\nu}\,\phi^{\kappa}\,_{\kappa}\phi^{\sigma}\,_{\sigma}\Big]=0 \,. 
\end{eqnarray} 

In the following we will assume that the background geometry is described 
by the flat Robertson-Walker metric $ds^{2}=-dt^{2}+a(t)^{2}d{\vec x}^{2}$, 
and we will derive the corrections to the Friedmann equations due to the 
presence of the coupled scalar field.  The  components of the  equations 
for the metric dynamics that are of interest for the cosmic evolution are 
the ones with $\mu=\nu=0$ (leading to the modified first Friedmann equation) 
and $\mu=\nu=i$ (leading to the modified second Friedmann equation).

To summarize the essential quantities: $g_{00}=g^{00}=-1$, 
$g_{ii}=a(t)^{2}=(g^{ii})^{-1}$ and the non-zero Christoffel symbols are 
\begin{eqnarray}
\Gamma^{i}\,_{0i}=\frac{\dot a}{a}\equiv H(t)\,,\quad\Gamma^{0}\,_{ii}=a\dot a 
\,. 
\end{eqnarray}
The non-zero components of the Riemann tensor are 
\begin{eqnarray}
R^{0}\,_{i0i}=a\ddot a\,,\quad R^{i}_{0i0}=-\frac{\ddot a}{a}\equiv\dot H+ 
H^{2}\,,\quad R^{i}\,_{jij}=\dot a^{2} 
\end{eqnarray}
while the ones for the Ricci tensor are 
\begin{equation}
R_{00}=-3\frac{\ddot a}{a}\,, \quad R_{ii}=a\ddot a+2\dot a^{2}
\end{equation} 
and the Ricci scalar is given by 
\begin{equation}
R=6\left[\frac{\ddot a}{a}+\left(\frac{\dot a}{a}\right)^{2}\right]\,. 
\end{equation} 

Taking all this into account we get the $\mu=\nu=0$ component of the 
metric evolution equation to be 
\begin{eqnarray}
&&3\frac{M_{P}^{2}}{2}\left(\frac{\dot a}{a}\right)^{2}-\frac{1}{4}\dot\phi^{2}-\frac{1}{4}a^{-2}\sum_{i}(\partial_{i}\phi)^{2}\nonumber\\
&&-\frac{c_{3}}{2M_{P}^{2}}\left[- \left(\frac{\dot a}{a^{2}}\right)^{2}\sum_{i}(\partial_{i}\phi)^{2}+9\left(\frac{\dot a}{a}\right)^{2}\dot \phi^{2}\right.\nonumber\\
&&\left.-a^{-4}\sum_{ij}(\partial_{i}\partial_{j}\phi)^{2}-4a^{-3}\dot a 
\dot\phi \sum_{i}\partial_{i}\partial_{i}\phi\right.\nonumber\\
&&\left.+a^{-4}\left(\sum_{i}\partial_{i}\partial_{i}\phi\right)^{2}\right]= 
\frac{1}{2}\rho_m \,, 
\end{eqnarray} 
where a dot denotes a standard time derivative, $\partial_{i}$ denotes the 
derivative with respect to the comoving spatial coordinate $i=\{x,y,z\}$, 
and $\rho_m$ is the energy density associated with the matter fields. 

The $\mu=\nu=i$ component of the equation for the metric evolution is 
\begin{eqnarray}
&&\frac{M_{P}^{2}}{2} (-2\ddot a a-\dot a^{2})-\frac{1}{2}\partial_{i}\phi\partial_{i}\phi+\frac{1}{4}a^{2}\left(-\dot\phi^2 +a^{-2}\sum_{j}(\partial_{j} 
\phi)^2\right)\nonumber\\
&&-\frac{c_{3}}{2 M_{P}^{2}}\Bigg[- \dot a^{2} \dot\phi^{2}-2a\ddot a 
\dot\phi^{2}-\left(\frac{\dot a}{a}\right)^{2} \sum_{j}(\partial_{j}\phi)^{2}+2(\partial_{i}\dot\phi)^{2} 
 \nonumber\\ 
&&+2\left(\frac{\ddot a}{a}+\left(\frac{\dot a}{a}\right)^{2}\right)(\partial_i \phi)^2-4\partial_{i}\dot\phi\frac{\dot a}{a}\partial_{i}\phi-2a^{-2}\sum_{j}(\partial_{j}\partial_{i}\phi)^{2}\nonumber\\
&&-2\frac{\dot a}{a} 
\dot\phi \partial_{i}\partial_{i}\phi-2\ddot\phi\partial_{i}\partial_{i}\phi-4\dot\phi \ddot\phi a\dot a+2(\partial_{i}\partial_{i}\phi)a^{-2}\sum_{j}(\partial_{j}\partial_{j}\phi)\nonumber\\
&&-2\sum_{j}\left[(\partial_{j}\dot\phi)^{2}-2\partial_{j}\dot\phi\frac{\dot a}{a}\partial_{j}\phi\right]+a^{-2}\sum_{j ,k}(\partial_{j}\partial_{k}\phi)^{2}\nonumber\\
&&+\left(2\frac{\dot a}{a}\dot\phi+2\ddot\phi\right) \sum_{j }\partial_{j}\partial_{j}\phi-a^{-2}\left(\sum_{j}\partial_{j}\partial_{j}\phi\right)^{2}\Bigg]\nonumber\\ 
&&=\frac{a^2P_m}{2} \,,\label{eq:EOM}
\end{eqnarray} 
where $P_m$ is the pressure of the matter fields. 

The equation for the field evolution is given by the variation of the 
action with respect to the field, requiring that $\delta S/\delta\phi=0$. 
So from the action of Eq.~(\ref{eq:fullaction}), and using 
$\nabla_{\mu}G^{\mu\nu}=0$, we get 
\begin{equation}
-g^{\mu\nu}\nabla_{\mu}\nabla_{\nu}\phi +\frac{2 c_{3}}{M_{P}^{2}}G^{\mu\nu}\nabla_{\mu}\nabla_{\nu}\phi=0 \,, 
\end{equation} 
as predicted in Eqs.~(\ref{eq:edef})--(\ref{eq:ekg}). 

Specializing to a scalar field smooth on the Hubble scale or below, 
as usually considered, we can neglect the spatial derivatives 
relative to the time derivatives and obtain the following modified 
Friedmann equations: 
\bea 
\frac{3M_P^2}{2}H^2&=&\frac{1}{2}\rho_m+\frac{1}{4}\dot\phi^2 
+\frac{9c_3}{2M_P^2}H^2\dot\phi^2\\ 
M_P^2\left(\frac{\ddot a}{a}+\frac{1}{2}H^2\right)&=&-\frac{1}{2}P_m 
-\frac{1}{4}\dot\phi^2 \nonumber\\
&&+\frac{c_3}{2 M_P^2} 
\left(H^2\dot\phi^2+2\frac{\ddot a}{a}\dot\phi^{2}
+4H\dot\phi\ddot\phi\right) \,. 
\eea 

In addition there is the equation of motion for the field $\phi$, 
the modified Klein-Gordon equation 
\be 
\bs 
0&=\ddot\phi\,\left(1+\frac{6c_3}{M_P^2}H^2\right)\\  
&\qquad + 3H\dot\phi \, 
\left[1+\frac{4c_3}{M_P^2}\left(\frac{\ddot a}{a}+\frac{1}{2}H^2\right)\right] \,, \label{eq:kg} 
\end{split} 
\ee 
although as usual only two of the three equations are independent.

\section{Cosmological Dynamics and Attractors} \label{sec:cos} 

Given the evolution equations we can now examine the cosmological 
dynamics and look for fixed point solutions or attractors for the 
dynamics insensitive to initial conditions.  The first Friedmann 
equation has a contribution modified from that of a canonical scalar 
field by a factor proportional to $C=c_3 H^2/M_P^2$.  So $C$ will be 
a key parameter.  We can in fact write the dynamics as an autonomous 
system of coupled first order differential equations 
\bea 
\frac{dC}{dN}&=&-C\,(1+w\op)\\ 
\frac{d\op}{dN}&=&-3w\op\,(1-\op) \,, 
\eea 
using the definition of the total equation of state 
$w_{\rm tot}\equiv -1-(1/3)d\ln H^2/dN=w\op$, where the second 
equality holds for the barotropic fluid having zero pressure, as 
for matter.  We will consider a matter plus scalar field universe 
for the following calculations. 

Examining the system of dynamical equations, we can immediately see 
several possible solutions of interest.  There may be fixed points 
at 1) $\op=0$, $C=0$, 2) $w=0$, $C=0$, 3) $\op=1$, $C=0$, and 
4) $\op=1$, $w=-1$.  We must examine the equations further to assess 
which are physical and stable, but the last in particular is of interest 
since it gives a cosmological constant solution despite having only 
kinetic terms in the action.  Recall there is no potential at all. 

The first Friedmann equation acts as a constraint, which we can write as 
\be 
1=\om+\frac{\dot\phi^2}{6M_P^2H^2}+\frac{3c_s\dot\phi^2}{M_P^4}\,. 
\ee 
The last two terms define the effective dark energy density 
\be 
\op=\frac{\dot\phi^2}{6M_P^2H^2}\,(1+18C)\,. \label{eq:ophiC} 
\ee 
To prevent ghosts we insist that the kinetic energy is non-negative, 
imposing the requirement 
\be 
C\ge-1/18\,. \label{eq:c18} 
\ee 

Using the second modified Friedmann equation we can define the effective 
pressure of the dark energy and hence its equation of state ratio 
$w=P_\phi/\rho_\phi$.  The result is 
\be 
w=\frac{1+30\,C+(P_m/\rho_\phi)\,6\op\,C(1-18C)}{1+(24-6\op)\,C+108(1+\op)\,C^2} \,, \label{eq:wC} 
\ee 
and as stated above we will take the barotropic pressure $P_m=0$.  We can 
investigate when it is possible to have $w=-1$, say.  There are two solutions, 
with $C=-1/6$ as mentioned earlier in Eq.~(\ref{eq:cdes}) and $C=-1/18$.  
The first violates the condition in 
Eq.~(\ref{eq:c18}) while the second one saturates it.  

Now let us return to examination of the four possible fixed points.  The 
first one corresponds to both the noncanonical nature, i.e.\ the kinetic 
coupling, and the dark energy as a whole fading away.  (The dark energy 
fading follows from the kinetic coupling vanishing since canonical kinetic 
energy redshifts away as $a^{-6}$.)  The second possible fixed point 
does not actually occur because $C=0$ in Eq.~(\ref{eq:wC}) implies 
$w=+1$, i.e.\ a (canonical) kinetic dominated evolution, not $w=0$.  
The third possibility also gives $w=1$, in which case $\op=1$ is only 
valid transiently unless there is no other component, so this is 
expected to be an unstable point. 

Case 4 has two branches: we have 
seen that the $C=-1/6$ root is invalid, leaving the $C=-1/18$ root. 
When $C=-1/18$, saturating the ghost protection condition, then by 
Eq.~(\ref{eq:ophiC}) the energy density $\op=0$ not 1, showing that this 
is not a fixed point.  However, if while $C$ is near $-1/18$ 
simultaneously $\dot\phi^2$ gets large, then a finite $\op$ is possible.  
This indeed turns out to be an unstable critical point, a saddle point 
with interesting 
quasi-attractor properties.  Thus we expect two main solutions: the 
dark energy fades away along with the kinetic coupling, but there may be 
an intermediate ``loitering'' phase with cosmological constant behavior 
dominating the universe. 

Because of Eq.~(\ref{eq:wC}) giving the relation $w(\op,C)$, and the 
first order nature of the coupled dynamical equations, if we specify 
$\op$ and $w$, say, at some particular time then the evolutionary 
trajectory is determined for all times.  We can define the present by 
$\op=0.72$ and generate a one parameter family of curves corresponding 
to different $w_0$, the value of the equation of state today, or 
different $C_0$.  

\begin{figure}[htbp!]
\includegraphics[width=\columnwidth]{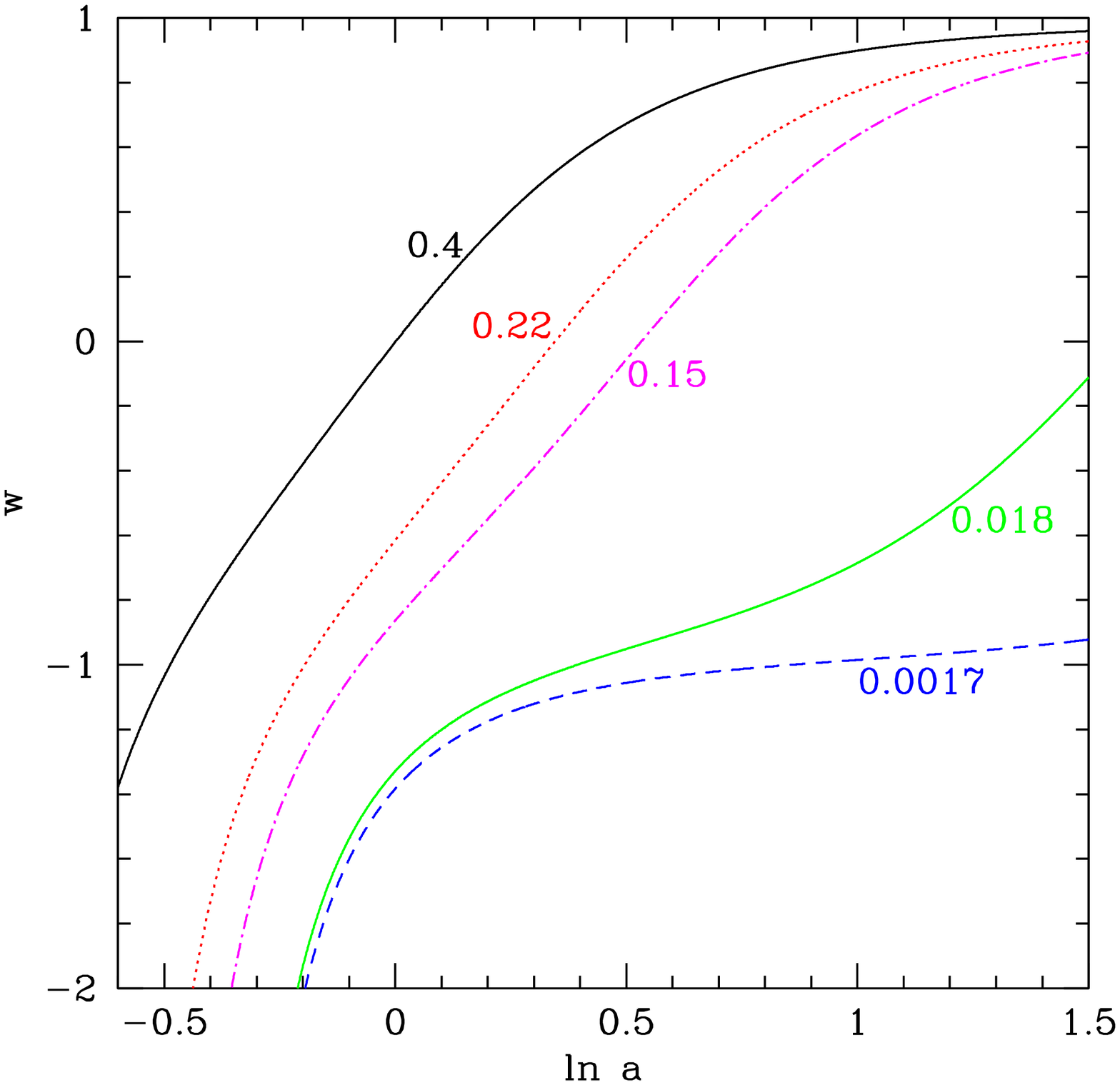}\\ 
\includegraphics[width=\columnwidth]{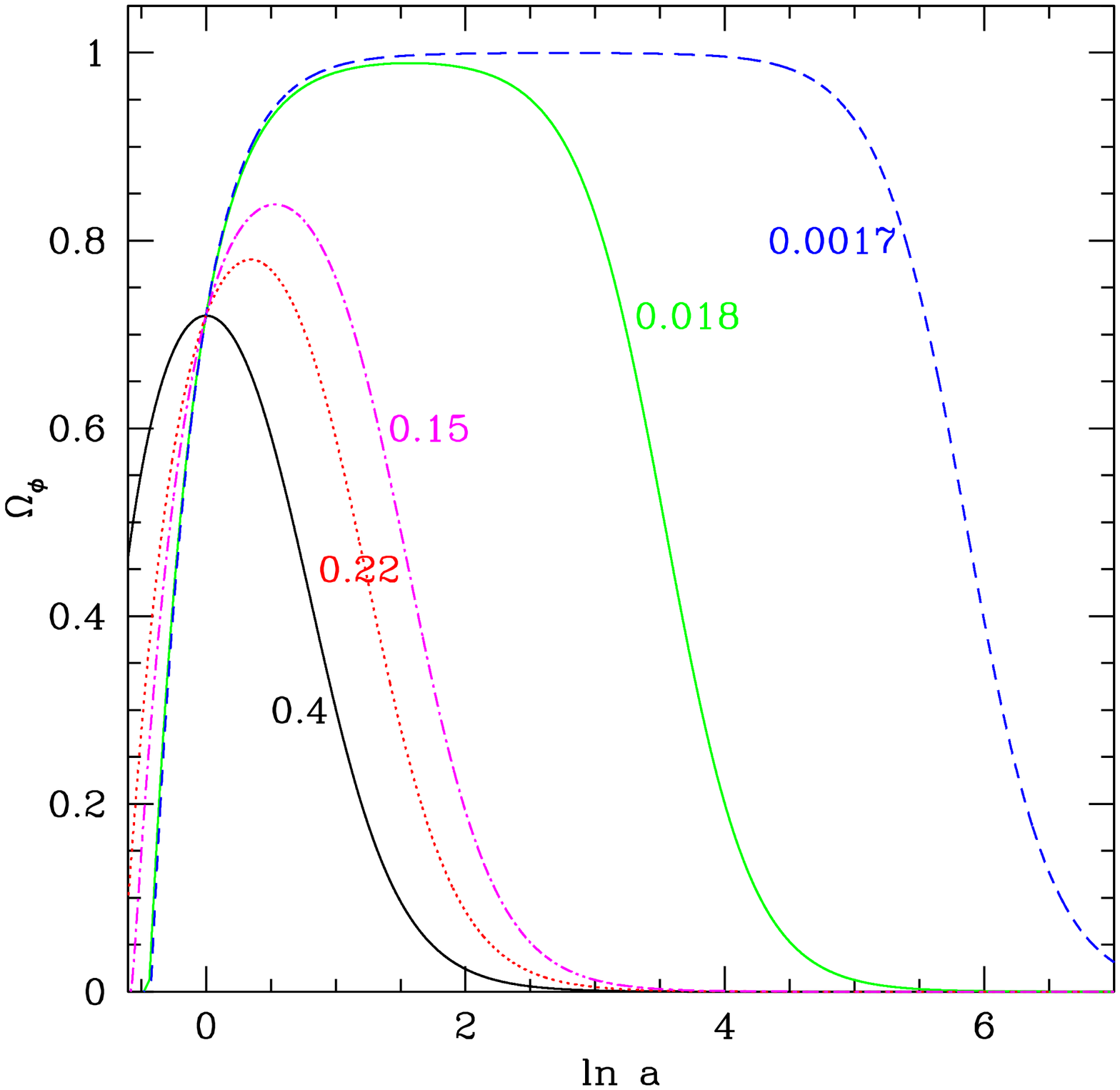} 
\caption{In the purely kinetic coupled gravity cosmology the evolution 
of the dark energy equation of state $w$ (top panel) is rapid, becoming 
relevant only at $z<1$.  From very negative values it then asymptotically 
approaches $w=+1$, though it can loiter near $w=-1$ for several e-folds 
of expansion if $\delta=(C_0-C_\star)/C_\star$ is small enough.  Curves 
are labeled with $\delta$.  During 
the loitering phase, the fractional dark energy density $\op$ (bottom 
panel) can approach 1, giving a cosmological constant dominated universe 
without a cosmological constant.  
}
\label{fig:dyn}
\end{figure}

Figure~\ref{fig:dyn} show the dynamical evolution of the purely kinetic 
coupled gravity cosmology for various values for $C_0$.  We define the 
fractional deviation $\delta=(C_0-C_\star)/C_\star$ away from the critical 
value $C_\star=-1/18$ as the key parameter.  As this gets small, the 
evolution of $w$ begins to loiter around the cosmological constant value 
$w=-1$.  The loitering can last for several e-folds of expansion as 
$\delta$ gets very small.  This certainly involves a fine tuning, but 
so does a true cosmological constant.

In the past the dark energy fades away very quickly, as $w\to -\infty$. 
This means that all early universe (in fact $z\gtrsim 1$) cosmology 
is unchanged from a standard matter (or earlier radiation) dominated 
universe.  In the 
asymptotic future, dark energy fades away as well, restoring a matter 
dominated universe.  Thus cosmic acceleration is a transient phenomenon 
in the purely kinetic coupled gravity scenario. 

We see that for moderate values of $\delta$, the evolution of $w$ is 
rapid and dark energy domination lasts for less than two e-folds of 
expansion; indeed for much of that time the dark energy does not act 
in an accelerating manner.  Large values of $\delta>0.4$ are not 
permitted because the universe never reaches a sufficient dark energy 
fraction $\op=0.72$ observed today.  However as $\delta$ gets small the 
quasi-stable attractor to $w=-1$ gets stronger and simultaneously the 
dark energy almost completes dominates the energy density for several 
e-folds.  We can thus get a period of cosmological constant dominated 
behavior, despite the absence of any cosmological constant (or potential 
of any sort), with only kinetic coupling terms. 

\begin{figure}[htbp!]
\includegraphics[width=\columnwidth]{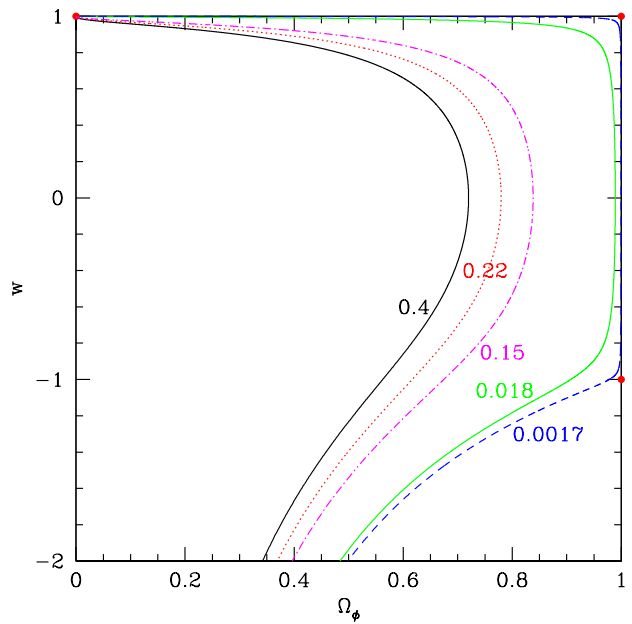}
\caption{In the phase space diagram for $\op$--$w$, one can clearly 
see the behavior of the trajectories near the three physical critical 
points (solid red dots).  The two critical points with $\op=1$ are 
unstable, but trajectories corresponding to 
$\delta=(C_0-C_\star)/C_\star\ll1$ (curves are labeled with $\delta$) 
can approach them closely, in fact loitering several e-folds near 
$w=-1$, $\op=1$.  Eventually, however, the dynamics leads to the 
stable attractor at $\op=0$, $w=1$ and cosmic acceleration ceases and 
dark energy fades away.  Note that curves with $\delta>0.4$ never 
achieve $\op=0.72$ and so do not describe our universe. 
}
\label{fig:opw}
\end{figure}

Figure~\ref{fig:opw} shows the $\op$--$w$ phase space, where the time 
coordinate (scale factor) runs along the trajectories.  We clearly see 
the cases with the small values of $\delta$ approach the quasistable 
fourth critical point at $(1,-1)$ but then depart (to see the length 
of the loitering one must look at Fig.~\ref{fig:dyn}).  These trajectories 
then move toward the unstable third critical point at $(1,1)$ but 
again are driven away, this time toward the stable first critical point 
at $(0,1)$ where dark energy has faded away.  For large enough $\delta$ 
though the trajectories immediately go to the stable critical point.

\section{Conclusions} \label{sec:concl}

We have considered the implications for the expansion history of the 
universe of modifying the standard general relativity action at the 
lowest possible order in the Planck mass (or equivalently in the Newton 
constant) through the addition of a coupling between functions of the 
metric and kinetic terms of a free scalar field.  This uses a purely 
kinetic approach, without any potential that would suffer high energy 
quantum corrections, and also obeys a shift symmetry. 

Starting from the most general form for the coupling we demonstrate that 
only one particular combination of all the possible terms is admissible 
to give no higher than second order field equations, 
if we impose the additional requirement of not having more than two 
derivatives of the field in the dynamical equations.  The resulting 
theory is characterized by only one parameter setting the strength of the 
coupling between the metric and the derivatives of the scalar field, and 
so has no more parameters than a cosmological constant model. 

The simple term involving the Einstein tensor and field kinetic term 
adds to Galileon gravity, often characteristic of higher dimension 
gravity theories, and also has ties to vector or Einstein aether gravity 
and disformal theories with a deformed metric. 

We have studied the evolution of the universe under the dynamics 
described by this action without a cosmological constant. The range 
of variation of the  free parameter of the theory is constrained on 
one side by the requirement of having positive kinetic energy of the 
field, and on the other side by the necessity  of producing the 
actually observed value of the dark energy density fraction today.  
The fine tuning is no worse than that of the cosmological constant. 
Despite the absence of a potential, a wide variety of dark energy 
equations of state is possible, ranging from phantom to stiff behaviors. 
We delineate and show the phase space evolution of the dynamics. 

Most interestingly, we find that for certain values of the one parameter 
the universe goes through a quasistable loitering phase that mimics a 
cosmological constant, without the necessity of adding any potential for 
the field.  At high redshift the universe has standard matter (and 
radiation) dominated behavior before the effective dark energy dominates.  
The loitering cosmological constant phase can last several e-folds before 
the effective dark energy fades away leaving the universe once again 
matter dominated. 


\acknowledgments

We thank Stephen Appleby for very helpful discussions and crosschecks.  
GG thanks the Institute for the Early Universe, Ewha University, for 
hospitality.  This work has been supported in part by the Director, 
Office of Science, Office of High Energy Physics, of the U.S.\ Department 
of Energy under Contract No.\ DE-AC02-05CH11231 and by World Class 
University grant R32-2009-000-10130-0 through the National Research 
Foundation, Ministry of Education, Science and Technology of Korea.


\end{document}